# A Short Wavelength GigaHertz Clocked Fiber-Optic Quantum Key Distribution System

Karen J. Gordon, Veronica Fernandez, Paul D. Townsend, and Gerald S. Buller

*Abstract*— **A quantum key distribution system has been developed, using standard telecommunications optical fiber, which is capable of operating at clock rates of greater than 1 GHz. The quantum key distribution system implements a polarization encoded version of the B92 protocol. The system employs vertical-cavity surface-emitting lasers with emission wavelengths of 850 nm as weak coherent light sources, and silicon single photon avalanche diodes as the single photon detectors. A distributed feedback laser of emission wavelength 1.3 μm, and a linear gain germanium avalanche photodiode was used to optically synchronize individual photons over the standard telecommunications fiber. The quantum key distribution system exhibited a quantum bit error rate of 1.4%, and an estimated net bit rate greater than 100,000 bits$^{-1}$ for a 4.2 km transmission range. For a 10 km fiber range a quantum bit error rate of 2.1%, and estimated net bit rate of greater than 7,000 bits$^{-1}$ was achieved.**

## I. INTRODUCTION

QUANTUM key distribution (QKD), first proposed by Bennett and Brassard in 1984 [1], is a means of distributing a *verifiably* secure key, between two or more users [2], over an unsecured channel [3]. A secure method of encryption – the "one-time pad" approach - was proposed in 1917 by Gilbert Vernam, and was proven absolutely secure by Claude Shannon in 1949 [4]. This encryption technique relies on a key that is, truly random, as long as the message itself, and used only once. Uniquely, QKD provides a method for distributing this type of key in a verifiably secure manner. QKD exploits the fundamental laws of quantum mechanics to achieve security; for example, the fact that the non-orthogonal polarization states of individual photons cannot be simultaneously measured with arbitrarily high accuracy nor without perturbation. Hence, if the random key data is encoded and transmitted using such states secure key sharing can be achieved, because an eavesdropper can only ever obtain partial information on the key and will inevitably generate a detectable disturbance on the quantum communication channel. Bennett and Brassard's original QKD protocol

This work was supported by the UK Engineering and Physical Sciences Research Council project GR/N12466, and the European Commission's Framework Five EQUIS project (IST-1999-11594).

K. J. Gordon (k.j.gordon@hw.ac.uk), V. Fernandez, and G. S. Buller are with the School of Engineering and Physical Sciences, Heriot-Watt University, Riccarton, Edinburgh, EH14 4AS, UK.

P.D. Townsend is with the Photonic Systems Group, Physics Department, University College Cork, Cork, Ireland.

(BB84) employs two incompatible pairs of conjugate quantum observables, for example circular and linear-polarization states, to encode the data [1]. In Bennett's simpler B92 protocol two non-orthogonal states are utilized, for example two non-orthogonal linear-polarization states [3].

To date, many groups have demonstrated the possibility of free-space [5]-[8] and optical fiber-based QKD systems [9]-[14]. There has been growing interest in QKD systems at lower operating wavelengths of ~850 nm [15]-[22] due to the superior performance of technologically mature silicon single photon avalanche diodes (SPAD's). Silicon SPAD's exhibit high efficiencies at wavelengths below 1000 nm [22]-[23]. However, they do not exhibit the deleterious effects of afterpulsing commonly associated with Ge and InGaAs-based SPAD's which are capable of high efficiency detection at wavelengths of greater than 1000 nm [24]-[25]. This detector afterpulsing phenomenon has limited the performance of fiber-based QKD systems operating at the low-loss fiber windows of 1.3 μm and 1.55 μm wavelength; leading to key exchange rates of only < 50 bits$^{-1}$ at distances of > 50 km. The fiber loss at 850 nm wavelength (~2.2 dBkm$^{-1}$) is far too high to achieve transmission distances of this magnitude. However, for short distance (~10 km) applications of QKD in, for example, campus- or metropolitan-scale networks short wavelength systems are likely to offer the highest key exchange rates due to the superior performance of silicon SPAD's. Standard telecommunications fiber is not single-mode at a wavelength of 850 nm and this might be thought to prevent the implementation of QKD over deployed fiber networks. However, one of the current authors has previously demonstrated that mode-selective launch techniques can be employed to alleviate this problem [18]. The QKD system detailed in this paper builds on this original work, but explores the possibilities of faster key transmission rates and the implementation of different protocols.

The so-called B92 protocol was chosen here since it is simpler to implement in practice than the BB84 protocol, as only two states are required instead of four. The polarization encoded version of B92 proceeds as follows for an idealized system. The transmitter "Alice" and the receiver "Bob" each generate an independent random bit sequence. Alice then transmits her random bit sequence to Bob using a clocked sequence of linearly-polarized, individual photons with polarization angles chosen according to her bit values as given by 0° ≡ 0 and 45° ≡ 1. In each time period Bob makes a polarization measurement on an incoming photon by orientating the transmission axis of his polarizer according to his bit value as given by -45° ≡ 0 and 90° ≡ 1. It can be seen that Bob will only detect a photon (with probability one half) in the time slots where his polarizer is not crossed with

Alice's. We refer to these instances as "unambiguous" since when they occur Alice and Bob can be sure that their polarization settings were not orthogonal and, consequently, that their bit values were the same (both 0 or both 1). Conversely, the instances in which Bob receives no photon are referred to as "ambiguous" since they can arise either from the cases where Alice and Bob's polarizers were crossed or from the cases where the polarizers were not crossed, but Bob failed (with probability one half) to detect a photon. Bob then uses an authenticated public channel to inform Alice of the time slots in which he obtained an unambiguous result (one in four on average) and they use the shared subset of their initial random bit sequences represented by these time slots as a key. The level of intervention by an eavesdropper "Eve" can then be quantified in the usual way by analyzing the error rate for the key exchange.

It is important to note that B92 is not as inherently secure as BB84, since Eve can in principle perform a potentially undetectable intercept-resend attack. With this type of attack, Eve chooses to only resend a photon to Bob when she obtains an unambiguous measurement outcome and hence knows Alice's polarization setting. This would not cause any depolarization-induced errors in the transmission, but would lead to a decreased photon arrival rate that would alert Bob to Eve's presence. However, if the quantum channel is lossy (as is the case with optical fiber) then Eve can in principle substitute a lower loss channel to compensate for her reduced photon transmission rate and hence avoid detection. Various approaches have been discussed to avoid this problem including the use of bright reference pulses in the original interferometric version of B92 [3]. We mention other defenses further below, but also note that the QKD system described here can be developed to implement the BB84 protocol with several straightforward changes. BB84 does not suffer from this security deficiency since, with four polarization states, Eve cannot be sure that she has determined the state of any given photon with deterministic accuracy.

We have demonstrated a QKD system operating up to a clock frequency of 1 GHz over standard telecommunications fiber lengths of up to 15 km by implementing the B92 protocol using linear-polarization encoding. Error correction and privacy amplification were not implemented in the work presented in this paper, thus the net bit rate after error correction and privacy amplification was estimated. The quantum bit error rate (QBER), raw bit rate, and estimated net bit rate (NBR) after error correction and privacy amplification for 4.2 km of fiber was 1.4%, ~300 kbits⁻¹, and ~130 kbits⁻¹ respectively. Correspondingly over 10 km of fiber, the QKD system demonstrated a QBER of 2.1%, raw bit rate of ~18 kbits⁻¹, and a NBR of ~7 kbits⁻¹. However, the NBR drops rapidly above 10 km due to the relatively high level of fiber loss. To the best of our knowledge, the work described in this paper represents the highest net bit transmission rate for any demonstration QKD system.

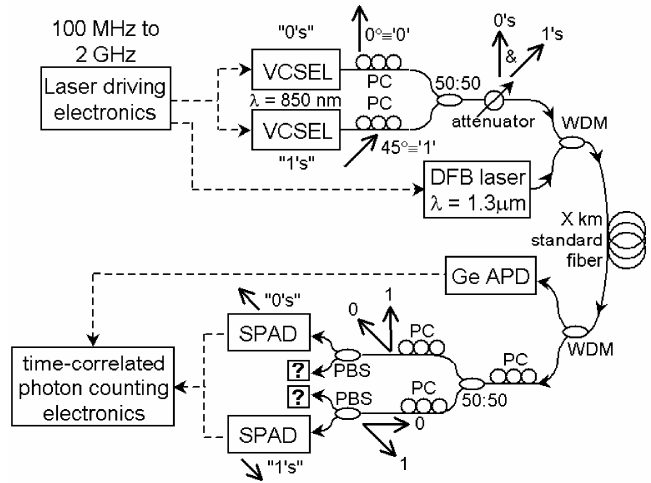

Fig. 1. Schematic diagram of the quantum key distribution experiment. PBS: Polarization splitter. PC: Polarization controller. WDM: Wavelength division multiplexor. APD: Avalanche photodiode. VCSEL: Vertical-cavity surface-emitting laser. SPAD: Single photon avalanche diode. DFB: Distributed feedback laser. The arrows represent polarization states of individual photons propagating in the optical fiber.

## II. EXPERIMENTAL

### A. Quantum key distribution system design

The characterized QKD system is shown in Fig. 1. The QKD system is entirely fiber-based, and an interchangeable length of standard telecommunications fiber separates Alice and Bob, the transmitter and receiver respectively. For characterization purposes, a non-return to zero (NRZ) pseudo-random bit sequence or a programmable bit stream was used to drive the laser source. Two 850 nm wavelength vertical-cavity surface-emitting lasers (VCSEL's) are used to implement the B92 protocol. In the case of the BB84 protocol, four VCSEL's could be used to implement the protocol, or one VCSEL could be used in conjunction with a suitable polarization modulator [26]. The light emitted from the VCSEL's is linearly-polarized and spatially and spectrally single mode when driven under the correct conditions. These VCSEL's typically exhibit a spectral line-width of ~0.1 nm under high-speed modulation, and are temperature tuned to have identical wavelengths, in order to avoid spectral interrogation by an eavesdropper. The VCSEL's were driven over a frequency range of 100 MHz to 2 GHz. A mean photon number (μ) of 0.1 photons per pulse was used for all experiments, meaning that the photon transmission rate from Alice varied between 10 Mbits⁻¹ and 200 Mbits⁻¹ depending on the chosen clock frequency.

The polarized light from each VCSEL was launched into 5.5 μm core diameter optical fiber, which was single mode at a wavelength of 850 nm. Both Alice and Bob utilized 850 nm single mode fiber, but were separated by an interchangeable length of standard telecommunications fiber. The standard telecommunications fiber supported more than one mode at an operating wavelength of 850nm. However, when a length of standard telecommunications fiber is fusion spliced to a fiber of smaller core diameter then the fundamental mode is excited and the second order modes are greatly suppressed. Thus, only one stable spatial mode propagates in the fiber at all times [18]. This method ensures that no significant temporal spreading or depolarization of the optical pulses due to modal dispersion occurs, even at Gbits$^{-1}$ clock frequencies.

The polarized light from each VCSEL once launched into the separate 5.5 μm core diameter single mode fibers are manipulated using fiber polarization controllers. The linear-polarization can be manipulated into two non-orthogonal polarization states (as required for the B92 protocol) at an angle of 45° with respect to each other as illustrated by the arrows in fig. 1. The VCSEL's are arranged so that one VCSEL represents a "0" and the other represents a "1". The non-orthogonal polarization states are combined by a polarization independent 50:50 coupler, and attenuated to a mean number of 0.1 photons per laser pulse (μ) using a programmable attenuator.

A distributed feedback (DFB) laser of emission wavelength 1.3 μm was used to optically clock and time-stamp the individual photons transmitted to Bob. The DFB laser was driven at a sub-multiple of the clock frequency denoted as the synchronization frequency. Where, the clock frequency was the driving frequency of the VCSEL's.

A germanium avalanche photodiode (APD) biased ~1 V below avalanche breakdown (in the linear multiplication regime) was used to detect the optical synchronization pulse and convert it into an electrical signal. Cross-talk of the 1.3 μm wavelength light into the 850 nm wavelength channel was reduced to the required level using a high-finesse band-pass filter centered at 850 nm.

The dual polarization states arriving at Bob after propagating through X km of standard telecommunications fiber, are manipulated to enable 25% of the transmitted data to be measured unambiguously. This is achieved by using two fiber based polarizing beam splitters (PBS's) and two silicon SPAD's. Since the driving signal of the VCSEL's was NRZ, the unambiguous data could not be recombined with a delay in one of the fiber channels and measured using only one SPAD, as described in other work [9], [11], [18]. We will denote the fiber channel from which the unambiguous "0's" are measured as channel 0, and correspondingly the channel from which the unambiguous "1's" are measured as channel 1.

Note that there are two other fiber ports from the polarizing beam splitters indicated in fig. 1 by a "?". These ports indicate the two fiber channels that contain the remaining 75% of the ambiguous data. The photons in these ports could be measured using another two SPAD's, and the collected data could be analyzed statistically to detect the presence of an eavesdropper [27]. An optimal method of unambiguously sifting the data transmitted by Alice could be employed, and was proposed in [28]. This optimal method yields ~29% of the photons incident at Bob's receiver to be measured unambiguously using a non-orthogonal separation angle of 45°. However, for the purposes of these characterization measurements a simpler method of using only two polarizing beam splitters is employed.

The silicon SPAD's used in these characterization experiments were Perkin Elmer SPCM-AQR-12's. These SPAD's were used in conjunction with a time-correlated photon-counting acquisition card and a personal computer to collect the transmitted data. These particular Si SPAD's had a detection efficiency of ~40% at the operating wavelength of 850 nm, and a dark count rate of ~180 counts$^{-1}$ each. These Si SPAD's typically have a jitter of between 300 and 400 ps.

In this configuration the QKD system had typical system insertion losses of ~17.0 dB. This loss includes the 75% loss caused by the unambiguous bit filtering technique at the PBS's and the ~40% detection efficiency of the Si SPAD's. However, this system loss does not include the extra transmission fiber losses of ~2.2 dBkm$^{-1}$. Several lengths of fiber were used to characterize the QBER and the NBR of the QKD system. When a required fiber length was not readily available, extra attenuation could be added using the attenuator in order to simulate the losses induced by a specific length of standard telecommunications fiber. This method has been shown to generate realistic measurement results in [11] and is generally applicable as long as there is no significant pulse spreading or pulse depolarization due to chromatic dispersion or polarization mode dispersion in the fiber. In this paper we show that the simulated fiber attenuation measurements correlate closely with measurements taken using actual lengths of standard telecommunications fiber in both QBER and NBR analyses.

In any communication system based on standard fiber the polarization state of the light at the receiver can drift with time due to thermally- or mechanically-induced variations in the fiber birefringence. Clearly this is an important issue for a polarization-encoded QKD system since the transmitter and receiver must establish and maintain a shared polarization reference in order to implement key sharing. Fortunately, in most deployed fiber systems this polarization drift is relatively slow and can in principle be compensated by means of automated polarization controllers (albeit at extra cost). In the current work measurements were taken over a relatively short time, typically between 60 seconds and 600 seconds, so the effects of polarization drift were minimized. In the future it would be desirable to integrate automated polarization controllers into the QKD system, in order to improve the stability and hence the practicality of the system.

### B. Data collection

The QKD test system described within this paper is under development, with no eavesdropper test, error correction, or privacy amplification. In order to characterize the QKD system a Becker & Hickl SPC-600 photon counting card was used to collect a series of repetitive data, using either, a pre-determined 8-bit repetitive word (10101010), or a $2^{15}$ - 1 (32,767) repetitive pseudo-random bit sequence. The data was collected in the form of a histogram of collected counts from each SPAD over a specific collection time (T), typically 60 seconds in the case of the 8-bit word, and 600 seconds in the case of the pseudo-random bit-sequence. An example of a recorded histogram is shown in fig. 2.

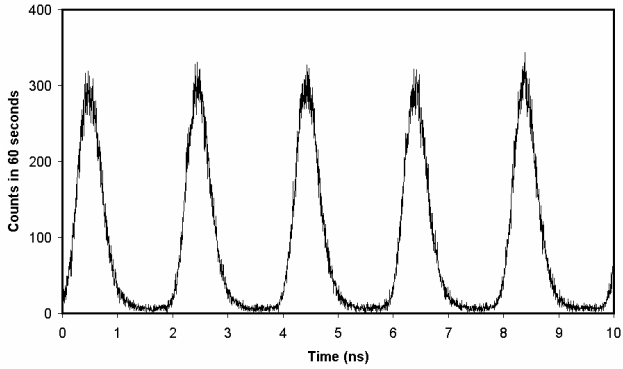

Fig. 2. A histogram collected over 60 seconds using the SPC-600 photon counting card in oscilloscope mode. The data was measured from channel 0, over a standard telecommunications fiber length of 11.07 km, using a repetitive 8-bit word of 10101010 at a clock frequency of 1 GHz.

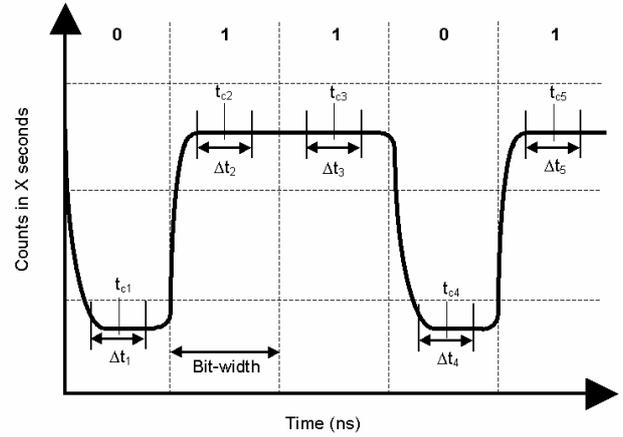

Fig. 3. Diagram illustrating each time period $\Delta t$ in which Bob accepts his the collected data, where $t_c$ is the center of each bit-period time window. $\Delta t$ can vary from 1% to 100% of the bit-width

The upper limit of the synchronization input of the SPC-600 photon counting card was 200 MHz. Thus, the synchronization frequency was a sub-multiple of the laser clock frequency in order for the synchronization frequency to be less than 200 MHz. The synchronization frequency was $1/16^{th}$ of the clock frequency in the case of a repetitive 8-bit word, and $1/(16 \times (2^{15} - 1))^{th}$ in the case of the pseudo-random bit-sequence.

For example, in the case of the repetitive 8-bit word, when operating at a clock frequency of 100 MHz, corresponding to a bit width (laser pulse width) of 10 ns, the synchronization frequency was divided to 6.25 MHz. Correspondingly, a clock frequency of 1 GHz, with a bit width of 1 ns, produced a synchronization frequency of 62.5 MHz. In the case of the pseudo-random bit-sequence, measurements were only taken at a clock frequency of 1 GHz with a corresponding synchronization frequency of 1.9 kHz.

In an installed QKD system operating at these data rates, the output from both SPAD's would be simultaneously collected and stored in the memory of the personal computer with the exact arrival time of each photon. This provides the information required by Alice and Bob in their post-transmission discussion for the discard of the irrelevant bits, and the subsequent eavesdropper test (QBER), error correction and privacy amplification. The volume of data required for accurate system characterization made this method difficult to implement, hence the technique of system analysis using repetitive input signals was adopted for the measurements described in this paper. Since the data transmitted between Alice and Bob is synchronized, Bob can choose to disregard photons that do not arrive in a specific time window ($\Delta t$) with respect to the synchronization pulse. For example, for a clock frequency of 1 GHz and bit width of 1 ns, Bob can choose to only accept photon counts which occur in a window $\Delta t = 50\%$ of the bit width, i.e. 0.5 ns. Since data at the change between each bit period introduces further error due to the temporal response of the each VCSEL and each SPAD. This process has been illustrated in fig. 3, where five bit periods are illustrated for channel 1 for a bit sequence of 01101. In this instance Bob accepts all of the data that is collected in the specific time intervals $\Delta t$ and disregards any counts outside these time periods. Bob can calculate when a detected count from a SPAD was outside the time window $\Delta t$ by referring to the synchronization pulse, giving the arrival time of each photon. From the collected data we can ascertain the QBER, raw bit rate, sifted bit rate, and estimate the net bit rate, which are defined in sections C and D.

## C. Raw bit rate, sifted bit rate and net bit rate

The raw bit rate per second for this QKD system can be defined as,

$$R_{raw} = n_s n_f \mu \nu \qquad (1)$$

Where $n_s$ is the system loss of the QKD system including the insertion losses at the polarizing beam splitters and detection efficiency of the SPAD's, $n_s$ was approximately 17.0 dB. $n_f$ is the transfer loss of the standard telecommunications fiber which is dependent on the length of fiber used, which has a loss of ~2.2 dBkm$^{-1}$ at a transmission wavelength of 850 nm. $\mu$ is the mean photon number per laser pulse, which was set to 0.1 photons per pulse for all measurements described in this paper. $\nu$ is the clock frequency of the system, i.e. the VCSEL driving frequency, where $\nu$ could be set to any frequency between 100 MHz and 2 GHz.

The sifted bit rate per second for this system will be defined as the proportion of the raw bit rate accepted by Bob in each time interval $\Delta t$. This sifted bit rate includes correct and incorrect counts, which are used to calculate the QBER, and the NBR. Hence, the sifted bit rate is dependent on Bob's chosen value of $\Delta t$.

$$R(\Delta t)_{sift} = \sum_{n=1}^{N} R(\Delta t)_{raw} \qquad (2)$$

Where $R(\Delta t)_{raw}$ is the sum of raw counts in one time interval $\Delta t$ averaged over one second, and summed over all bit periods giving $R(\Delta t)_{sift}$. $R(\Delta t)_{sift}$ is the bit rate per second prior to error correction and privacy amplification.

The NBR ($R(\Delta t)_{net}$) per second after error correction and privacy amplification can be approximated using (3) [10], [29], [30].

$$R(\Delta t)_{net} \approx \left[ 1 + Q\log_2 Q - \frac{7}{2}Q - I_{AE}\left(1 - (1-Q)\log_2(1-Q) - \frac{7}{2}Q\right) \right] R(\Delta t)_{sift} \qquad (3)$$

Where Q is the QBER, which is dependent on $\Delta t$, and $I_{AE}$ is the amount of information an eavesdropper could gain.

$$I_{AE(max)} = 1 - \cos\theta \qquad (4)$$

For these approximations, we assume the *worst case scenario*, i.e. that the maximum unambiguous information is attained by an eavesdropper. Where θ is the angle between the non-orthogonal polarization states, thus when θ = 45° then the maximum information gain by an eavesdropper is approximately 29% [28], as shown in (4). Assuming the maximum information gain by an eavesdropper means the approximated value of $R_{net}$ is at its minimum value. $R(\Delta t)_{net}$ is dependent on the QBER and $R(\Delta t)_{sift}$, which is dependent on Bob's chosen value of $\Delta t$.

### D.  Quantum bit error rate

The QBER is the ratio of the total number of incorrect counts i.e. measuring a 0 instead of a 1 expressed as a percentage of the total number of correct and incorrect counts ($R(\Delta t)_{sift}$) (5). Three main factors contribute towards the detected incorrect counts.  (i) The dark count rate of the SPAD's.  (ii) Polarization leakage caused by the imperfect polarization extinction ratio of the PBS's.  (iii) A combination of the temporal response of the VCSEL's and the SPAD's which can cause a count to be detected in the incorrect time window.  At longer fiber transmission lengths, chromatic dispersion will also cause temporal broadening of the detected pulse shape, in turn adversely affecting the QBER.  At a wavelength of 850 nm temporal broadening due to chromatic dispersion is ~100 ps/nm.km in standard telecommunications fiber.  The spectral line-width of the VCSEL's under high-speed modulation is ≤ 0.1 nm, hence over 10 km of fiber the temporal broadening of the pulse due to chromatic dispersion is ≤ 100 ps.  The VCSEL laser pulse width at a clock frequency of 1 GHz is 1 ns and the temporal response of the Si SPAD's is ~350 ps.  Hence, the detectable error contribution from chromatic dispersion at longer fiber lengths is low.  However, at frequencies greater than 1 GHz, and SPAD's with a temporal response << 300 ps the error due to chromatic dispersion will have a more pronounced effect on the QBER.  The QBER in simple terms is defined in (5).

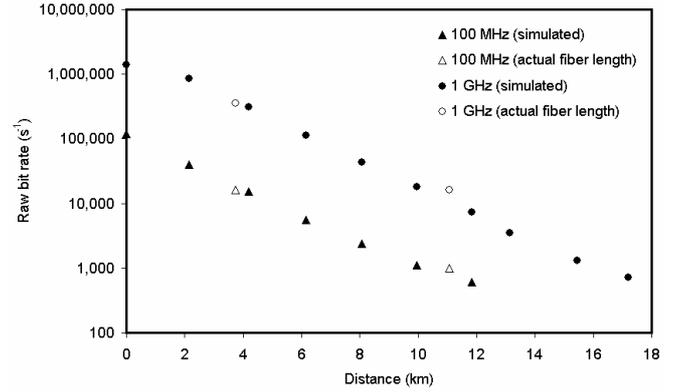

Fig. 4.  A logarithmic graph showing the $R_{raw}$ versus fiber length for the clock frequencies of 100 MHz and 1 GHz, and μ = 0.1 photons per pulse.  Where the *simulated* points are fiber lengths simulated using extra attenuation.

$$QBER = \frac{C_{incorrect}}{C_{correct} + C_{incorrect}} \qquad (5)$$

Where $C_{correct}$, is the number of correct counts, and $C_{incorrect}$ is the number of incorrect counts per unit of time.  For this QKD system the QBER calculation is defined as:

$$QBER(\Delta t) = \frac{\sum_{n=1}^{N} C(\Delta t)_{incorrect}}{\sum_{n=1}^{N} C(\Delta t)_{correct} + \sum_{n=1}^{N} C(\Delta t)_{incorrect}} \qquad (6)$$

Where $C(\Delta t)_{incorrect}$ is the number of incorrect counts in the time period $\Delta t$, summed over all bit periods N in which a photon should not be detected in either channel 0 and 1.  Correspondingly $C(\Delta t)_{correct}$, is the number of correct counts in the time period $\Delta t$, summed over all bit periods N in which a photon is correctly detected.  Where $\Delta t$ is a pre-determined percentage of the bit period time window.  For example at 1 GHz the bit period is 1 ns, therefore a $\Delta t$ = 50% (0.5 ns) could be chosen by Bob as valid data and anything outside this time period is discarded.

In order to characterize the QKD system, repetitive data from the VCSEL's was collected for a pre-determined period of time.  The photon count distributions were then analyzed to ascertain the QBER, sifted bit rate and to estimate NBR for specific lengths of fiber.  The raw bit rate was measured by either using the SPC-600 photon counting card or a photon counter, which counts the number of detected counts in a specified time interval.

The QBER of the QKD system has three main contributing factors as discussed earlier, these can be expressed as,

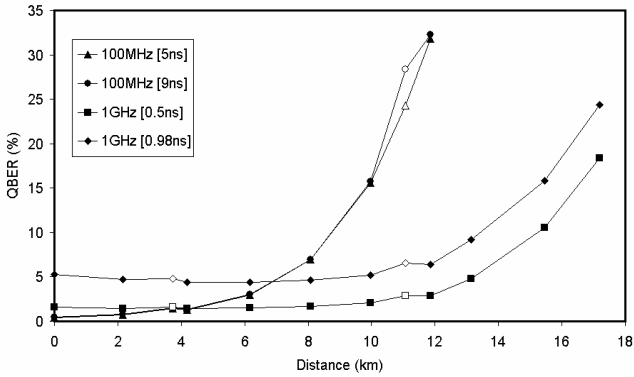

Fig. 5.  A logarithmic graph showing the QBER versus fiber length for the clock frequencies of 100 MHz and 1 GHz.  As in Fig. 4 the points taken with actual lengths of standard telecommunications fiber are highlighted in white.

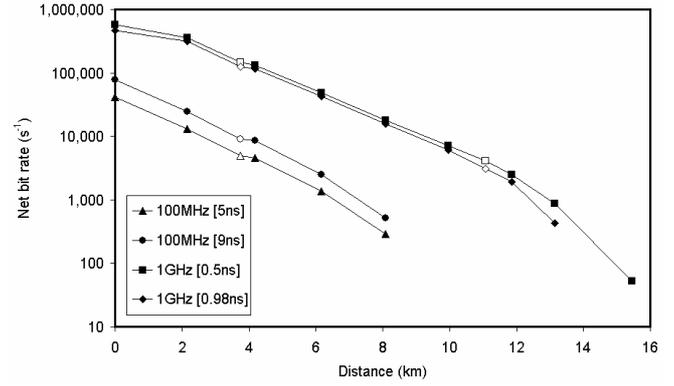

Fig. 6.  A logarithmic graph showing the NBR versus fiber length for the clock frequencies of 100 MHz and 1 GHz.  As in Fig. 4 the points taken with actual lengths of standard telecommunications fiber are highlighted in white.

TABLE I

RAW BIT RATE, SIFTED BIT RATE, NET BIT RATE, AND QUANTUM BIT ERROR RATE AT A CLOCK FREQUENCY OF 100 MHz.

| Distance (km) | $R_{raw}$ (bits$^{-1}$) | $R_{sift}$ (bits$^{-1}$) $\Delta t$=5ns | $R_{sift}$ (bits$^{-1}$) $\Delta t$=9ns | $R_{net}$ (bits$^{-1}$) $\Delta t$=5ns | $R_{net}$ (bits$^{-1}$) $\Delta t$=9ns | QBER (%) $\Delta t$=5ns | QBER (%) $\Delta t$=9ns |
|---|---|---|---|---|---|---|---|
| 0 | 119,257 | 61,948 | 117,698 | 41,147 | 77,671 | 0.4 | 0.4 |
| 2.15 | 39,450 | 20,419 | 38,675 | 12,930 | 24,329 | 0.7 | 0.8 |
| **3.75** | **16,298** | **8,569** | **15,839** | **4,971** | **9,073** | **1.4** | **1.5** |
| 4.19 | 15,122 | 7,805 | 14,807 | 4,583 | 8,639 | 1.3 | 1.4 |
| 6.16 | 5,595 | 2,882 | 5,361 | 1,357 | 2,505 | 3.0 | 3.0 |
| 8.08 | 2,401 | 1,220 | 2,192 | 293 | 520 | 6.9 | 7.0 |
| 9.96 | 1,109 | 554 | 917 | / | / | 15.6 | 15.7 |
| **11.07** | **990** | **493** | **717** | **/** | **/** | **24.3** | **28.4** |
| 11.85 | 611 | 291 | 425 | / | / | 31.8 | 32.3 |

$$\mathrm{QBER}(\Delta t) = \frac{R(\Delta t)_{leak} + R(\Delta t)_{dark} + R(\Delta t)_v}{R(\Delta t)_{sift}}. \quad (7)$$

Where $R(\Delta t)_{leak}$ is the contribution to the error due to the incorrect counts in each SPAD channel in each time window, $\Delta t$, caused by polarization leakage of the incorrect polarization state at each polarizing beam splitter.  $R(\Delta t)_{dark}$ is the error component caused by the incorrect counts in each SPAD channel in each time window, $\Delta t$, caused by the dark counts generated by each SPAD.  Dark counts are caused by thermally generated carriers initiating an avalanche, which is registered as a false event.  Such false events can also be caused by after-pulsing caused by carriers being released from trap centers some time after an avalanche event, however these effects are generally less evident in state-of-the-art silicon SPAD's operated near room temperature.  $R(\Delta t)_v$ are the photon counts contributing to the error caused by the temporal response of both the VCSEL's and the SPAD's.  This error contribution is low at lower frequencies such as 100 MHz, however $R(\Delta t)_v$ increases rapidly as $v$ is increased from 100 MHz to 2 GHz.

TABLE II

RAW BIT RATE, SIFTED BIT RATE, NET BIT RATE, AND QUANTUM BIT ERROR RATE AT A CLOCK FREQUENCY OF 1 GHz.

| Distance (km) | $R_{raw}$ (bits$^{-1}$) | $R_{sift}$ (bits$^{-1}$) $\Delta t$=0.5ns | $R_{sift}$ (bits$^{-1}$) $\Delta t$=0.98ns | $R_{net}$ (bits$^{-1}$) $\Delta t$=0.5ns | $R_{net}$ (bits$^{-1}$) $\Delta t$=0.98ns | QBER (%) $\Delta t$=0.5ns | QBER (%) $\Delta t$=0.98ns |
|---|---|---|---|---|---|---|---|
| 0.00 | 1,415,588 | 1,011,117 | 1,409,336 | 570,123 | 465,665 | 1.6 | 5.3 |
| 2.15 | 856,095 | 623,867 | 853,987 | 360,616 | 311,190 | 1.4 | 4.7 |
| **3.75** | **354,372** | **260,216** | **353,866** | **147,660** | **126,302** | **1.6** | **4.8** |
| 4.19 | 307,738 | 228,251 | 307,103 | 132,350 | 116,868 | 1.4 | 4.4 |
| 6.16 | 112,844 | 84,237 | 112,565 | 48,366 | 42,799 | 1.5 | 4.4 |
| 8.08 | 43,282 | 32,299 | 43,128 | 18,076 | 15,824 | 1.7 | 4.7 |
| 9.96 | 18,278 | 13,363 | 18,174 | 7,111 | 6,062 | 2.1 | 5.2 |
| **11.07** | **16,277** | **8,502** | **11,679** | **4,073** | **3,045** | **2.8** | **6.6** |
| 11.85 | 7,231 | 5,223 | 7,145 | 2,494 | 1,908 | 2.9 | 6.4 |
| 13.15 | 3,494 | 2,467 | 3,429 | 882 | 432 | 4.8 | 9.2 |
| 15.45 | 1,317 | 865 | 1,265 | 53 | / | 10.6 | 15.8 |
| 17.20 | 725 | 435 | 682 | / | / | 18.4 | 24.3 |

### E. Experimental results at 100 MHz and 1 GHz using a repetitive 8-bit word

In this section, we report a selection of results taken at the clock frequencies of 100 MHz and 1 GHz using an 8-bit repetitive bit sequence of 10101010 to drive the VCSEL's. We report the raw bit rate, sifted bit rate, net bit rate, and QBER measured for both clock frequencies.  Fig. 4 shows $R_{raw}$ plotted against fiber length for the clock frequencies of 100 MHz and 1 GHz.  $R_{raw}$ is the same whether the driving signal is an 8-bit repetitive bit-sequence or a $2^{15}$ -1 pseudo-random bit sequence.  $R_{sift}$, and QBER are dependent on the value of $\Delta t$ chosen by Bob.  For 100 MHz two values of $\Delta t$ were chosen 50% and 90%, 5 ns and 9 ns respectively, where the bit width was 10 ns.  At 1 GHz two values of $\Delta t$ were chosen, 50%, and 98%, giving 0.5 ns, and 0.98 ns respectively, where the bit width was 1 ns.  In fact, many values of $\Delta t$ were calculated for 1 GHz, but only two are shown for clarity.  The values obtained experimentally at 100 MHz and 1 GHz are shown in Table I and Table II respectively.  Fig. 5 shows a plot of the QBER expressed as a percentage versus fiber length, and Fig. 6 shows a plot of the estimated NBR versus fiber length, for both 100 MHz and 1 GHz clock frequencies.

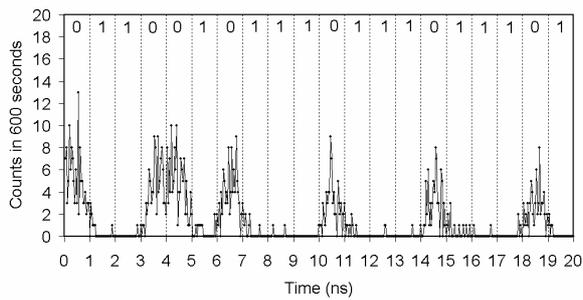

Fig. 7. Collected data for channel 0, taken over 600 seconds, showing the first 20 bits of the 127 bits of the $2^{15}$ - 1 pseudo-random bit-sequence collected and analyzed for a fiber length of 3.75 km.

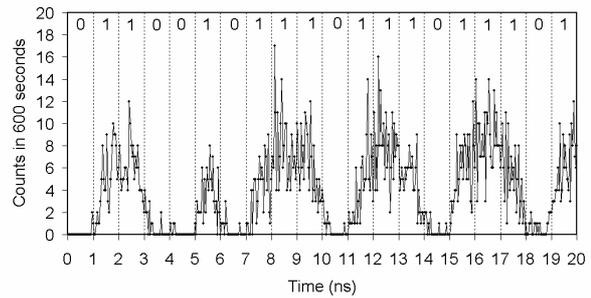

Fig. 8. Collected data for channel 1, taken over 600 seconds, showing the first 20 bits of the 127 bits of the $2^{15}$ - 1 pseudo-random bit-sequence collected and analyzed for a fiber length of 3.75 km.

TABLE III

COMPARISON QUANTUM BIT ERROR RATE AT A CLOCK FREQUENCY OF 1 GHz FOR AN 8-BIT REPETITIVE WORD 10101010 AND A 127-BIT PSEUDO-RANDOM BIT SEQUENCE.

|  | 0 km | 3.75 km | 6.01 km |
|---|---|---|---|
| Pseudo ($\Delta t$=0.5ns) | 2.7 | 1.6 | 3.0 |
| 8-bit word ($\Delta t$=0.5ns) | 1.6 | 1.6 | 1.5 |
| Pseudo ($\Delta t$=0.98ns) | 5.1 | 3.8 | 6.3 |
| 8-bit word ($\Delta t$=0.98ns) | 5.3 | 4.8 | 4.4 |

TABLE IV

COMPARISON NET BIT RATE AT A CLOCK FREQUENCY OF 1 GHz FOR AN 8-BIT REPETITIVE WORD 10101010 AND A 127-BIT PSEUDO-RANDOM BIT SEQUENCE.

|  | 0 km | 3.75 km | 6.01 km |
|---|---|---|---|
| Pseudo ($\Delta t$=0.5ns) | 359,632 | 101,428 | 29,693 |
| 8-bit word ($\Delta t$=0.5ns) | 570,123 | 147,660 | 48,366 |
| Pseudo ($\Delta t$=0.98ns) | 474,129 | 139,432 | 31,653 |
| 8-bit word ($\Delta t$=0.98ns) | 465,665 | 126,302 | 42,799 |

At 100 MHz the bit rate is $1/10^{th}$ of that at 1 GHz. Hence, the QBER reaches insecure levels at shorter distances because the number of received photons from Alice is correspondingly less, in turn increasing the QBER according to equation (7). However, at 1 GHz there is a larger QBER at shorter distances (i.e. < 3 km) due to the error contribution of $R(\Delta t)_v$. This is due to the relatively slower temporal response of the VCSEL's and SPAD's with respect to the bit period, causing the detection of arriving photons to spread into adjacent time windows. This static offset error at 1 GHz can be reduced further by reducing $\Delta t$ below 0.5 ns, however this will also reduce the NBR. It was found that the optimal value of $\Delta t$, by varying $\Delta t$ between 0.1 ns and 0.98 ns at 1 GHz, was ~0.5 ns, giving the optimal combination of QBER and NBR. If $\Delta t$ is greater than 0.5 ns, the QBER is higher resulting in more data being discarded in the error correction and privacy amplification estimation, which results in a lower NBR. Conversely, if $\Delta t$ is less than 0.5 ns then the reduction in QBER does not outweigh the reduction in NBR, causing a significant reduction in NBR.

At a clock frequency of 100 MHz there is little difference between $\Delta t$ = 5 ns and 9 ns for a clock frequency of 100 MHz. This is due to relatively low temporal spreading of the detected pulses with respect to the bit width, from the response of both the VCSEL's and SPAD's. Hence, the QBER results for $\Delta t$ = 5 ns and 9 ns at 100 MHz are very similar.

*F. Experimental results at 1 GHz using a repetitive $2^{15}$ - 1 pseudo-random bit-sequence*

The results discussed in section E only give an approximation of the QKD system performance since the collected data stream was a repetitive 8-bit sequence of 10101010, and not a random bit sequence. Therefore, a more realistic approach in characterizing the QKD system is to drive the VCSEL's with the pseudo-random bit-sequence discussed earlier. Due to the nature of the data collection method, $2^6 - 1$ bits of the pseudo-random bit sequence were collected using the SPC-600 photon counting card and then analyzed. Since $2^6$ - 1 is only a small percentage of the $2^{15}$ - 1 bit-sequence, the number of counts collected by the photon counting card in this time was very low in comparison to the data collection method described in section E. Consequently, the data necessarily had to be collected for 600 seconds.

The pseudo-random measurements were taken at 0 km and over two lengths of fiber, 3.75 km, and 6.01 km. The resultant values of QBER and $R_{net}$ are compared in Table III and Table IV to the collected values for the repetitive 8-bit word given in section E. Fig. 7 and Fig. 8 give an example of a section of the complementary histograms, for a fiber length of 3.75 km. Both figures show the first 20 bits of the 127-bit sequence collected and analyzed. In this instance, the first twenty bits of the sequence were 01100,10111,01110,11101 and the border between each bit period is indicated by the dashed x-axis gridlines.

As discussed previously the pseudo-random measurements were collected for a period of 600 seconds, and since automated polarization controllers were not employed, the polarization within the fiber was subject to some polarization drift within the collection time. Consequently, there are some differences between measurements taken with a repetitive signal for which the collection time was 60 seconds, and the pseudo-random bit-sequence, as observed in Table III and Table IV. At fiber lengths greater than 6 km the number of collected counts in 600 seconds was too low to make an accurate estimation of the QBER and in turn the net bit rate. This was due to the collection method of the data, limiting the number of possible measurements using the pseudo-random bit sequence.

By comparing Fig. 2, Fig. 7, and Fig. 8, it is clear that there are differences between the collection methods described in sections E and F. In fact, the pseudo-random measurements give a better estimation of the NBR than using the 8 bit repetitive bit-sequence. More bits of data are discarded by Bob in the post-transmission data processing when a pseudo-random sequence compared to the 8-bit repetitive signal, when using the same value of Δt. The driving signal of the VCSEL's is NRZ. Therefore when one VCSEL is lasing for several bit periods at a time, i.e. several sequential 0's or 1's, then the probability distribution of detected photons has an average height in each bit period until the bit value changes. This is assuming of course that the laser pulse has a constant height until it is driven below threshold current. This effect can be observed in Fig. 7 and Fig. 8. Thus, when Bob uses a small Δt, for example 50%, and a random bit sequence he discards many useful photon counts between each Δt. This does not occur when collecting data using the 8-bit 10101010 method as more photons are statistically in the center of each peak (center of the period Δt). Hence, the 8-bit 10101010 method gives a high estimation of the sifted bit rate and NBR in comparison to the pseudo-random method. This difference is more applicable to smaller values of Δt, such as 0.5 ns, as is observed in Table IV. However, as Δt increases towards the full bit width, the sifted bit rate and estimated NBR calculated using both methods converge since only a very small percentage of the data is discarded in each case, see Table IV.

## III. CONCLUSIONS

This paper discusses a prototype quantum cryptography system, which has been used to demonstrate the feasibility of high bit rate QKD over *standard* telecommunications fiber at an operating wavelength of 850 nm. The experimental system is capable of operation at GHz clock rates, in order to exploit the high potential bit rate performance of silicon SPADs, and employs mode-selective launch techniques to prevent modal instabilities or dispersion in the fiber. We believe that this approach currently offers the highest potential key distribution rates for short distance (~10 km) applications of QKD in, for example, campus- or metropolitan-scale networks. The net bit rate obtained from the experimental system after error correction and privacy amplification was estimated to be greater than 0.5 Mbits[-1] for 0 km, 100 kbits[-1] for 4.2 km, and 7 kbits[-1] for 10 km of standard telecommunications fiber. These distances generated QBER's of 1.6%, 1.4%, and 2.1% respectively at a clock frequency of 1 GHz. Two methods of data collection were used in order to characterize the QKD system. It was found that both the 8-bit repetitive bit sequence and pseudo-random bit sequence measurements gave comparable QBER values. However, it was shown that the 8-bit repetitive bit sequence method gave slightly higher estimated values of NBR in comparison to the pseudo-random bit sequence measurements. The pseudo-random method was limited by the low number of collected counts using this data collection method.

Data communications VCSEL's capable of operating at greater than 1 GHz performance, and silicon SPAD's with a jitter of much less than 300 ps, are currently under investigation in order to improve the performance of the reported QKD system. Preliminary indications are that these improved components will enable the system to operate at clock frequencies of greater than 1 GHz with an improved QBER and estimated net bit rate, thus realizing the potential of net bit rates greater than Mbits[-1] over local area networks (LAN's) of a few kilometers.

### ACKNOWLEDGMENT

The authors gratefully acknowledge many useful discussions with Professor Sergio Cova (Politecnico di Milano) and Dr Rebecca Wilson (QinetiQ, Malvern). This work was supported by the UK Engineering and Physical Sciences Research Council project GR/N12466, and the European Commission's Framework Five EQUIS project (IST-1999-11594).